# When Autonomy Breaks: The Hidden Existential Risk of AI

Joshua Krook, University of Antwerp

**Abstract:**

AI risks are typically framed around physical threats to humanity, a loss of control or an accidental error causing humanity's extinction. However, I argue in line with the gradual disempowerment thesis, that there is an underappreciated risk in the slow and irrevocable decline of human autonomy. As AI starts to outcompete humans in various areas of life, a tipping point will be reached where it no longer makes sense to rely on human decision-making, creativity, social care or even leadership.

What may follow is a process of gradual de-skilling, where we lose skills that we currently take for granted. Traditionally, it is argued that AI will gain human skills over time, and that these skills are innate and immutable in humans. By contrast, I argue that humans may lose such skills as critical thinking, decision-making and even social care in an AGI world. The biggest threat to humanity is therefore not that machines will become more like humans, but that humans will become more like machines.

**Statements and Declarations**

All authors certify that they have no affiliations with or involvement in any organization or entity with any financial interest or non-financial interest in the subject matter or materials discussed in this manuscript.

# Introduction

AI risks are typically framed around physical threats to humanity, a loss of control or an accidental error causing humanity's extinction.[1] However, I argue in line with the gradual disempowerment thesis,[2] that there is an underappreciated risk in the slow and irrevocable decline of human autonomy. As AI starts to outcompete humans in various areas of life, a tipping point will be reached where it no longer makes sense to rely on human decision-making, creativity, social care or even leadership.

What may follow is a process of gradual de-skilling, where we lose skills that we currently take for granted. Traditionally, it is argued that AI will gain human skills over time,[3] and that these skills are innate and immutable in humans.[4] By contrast, I argue that humans may lose such skills as critical thinking, decision-making and even social care in an AGI world.[5] The biggest threat to humanity is therefore not that machines will become more like humans, but that humans will become more like machines.

As control of our society slips from our grasp, the very nature of what it means to be a human will shift to match this new equilibrium. Companies will hire AI workers, CEOs will outsource decisions to AI agents and everyday decisions, from what to eat today to what to wear to collect our UBI dollars, will be made by AI assistants (who claim to curate, but actually control, our lives). Far from orchestrating technology, we will be orchestrated *by* that technology,[6] with it deciding everything from what jobs we should do to who we should date, because it will literally *know better than us*.[7]

The more we outsource our decisions, the more we will suffer de-skilling,[8] where our capacity to make decisions for ourselves, be creative, and even provide support to others, is eroded. Once AGI is significantly smarter than humans, it becomes illogical to decide for ourselves what to do with our lives, only to achieve worse outcomes.[9] Detached from other people, unable to make decisions, unemployed and vulnerable to manipulation, humans will become lesser.

Unlike a traditional gradual disempowerment thesis,[10] I do not argue this is a human alignment problem. The AGI imagined in this paper would act in the best interests of humanity. Despite this, humans will lose autonomy. Analogies exist in our society, where conservatorships take agency away from individuals who cannot make decisions in their own best interests, and are therefore put under the care of a guardian.[11] In an imagined future, AGI would place the entire human race under a conservatorship for our own good, making smarter, faster, more rational decisions than we would ever be capable of making. This would place humans in a

---

[1] Hilliard, A., Kazim, E. & Ledain, S. Are the robots taking over? On AI and perceived existential risk. *AI Ethics* (2024). https://doi.org/10.1007/s43681-024-00600-9

[2] Kulveit, Jan, Raymond Douglas, Nora Ammann, Deger Turan, David Krueger, and David Duvenaud. "Gradual Disempowerment: Systemic Existential Risks from Incremental AI Development." arXiv, January 29, 2025. https://doi.org/10.48550/arXiv.2501.16946.

[3] Rainie, Janna Anderson and Lee. "3. Improvements Ahead: How Humans and AI Might Evolve Together in the next Decade." *Pew Research Center* (blog), December 10, 2018. https://www.pewresearch.org/internet/2018/12/10/improvements-ahead-how-humans-and-ai-might-evolve-together-in-the-next-decade/.

[4] World Economic Forum. "How We Can Elevate Uniquely Human Skills in the Age of AI," January 20, 2025. https://www.weforum.org/stories/2025/01/elevating-uniquely-human-skills-in-the-age-of-ai/.

[5] See similar arguments by:
Ahmad, S.F., Han, H., Alam, M.M. *et al.* Impact of artificial intelligence on human loss in decision making, laziness and safety in education. *Humanit Soc Sci Commun* **10**, 311 (2023). https://doi.org/10.1057/s41599-023-01787-8; Zhai, C., Wibowo, S. & Li, L.D. The effects of over-reliance on AI dialogue systems on students' cognitive abilities: a systematic review. *Smart Learn. Environ.* **11**, 28 (2024). https://doi.org/10.1186/s40561-024-00316-7

[6] Han, Byung-Chul, *Non-things: Upheaval in the Lifeworld*. (1st ed.) Polity. Cambridge, UK, 2022.

[7] Buttazzo G. Rise of artificial general intelligence: risks and opportunities. Front Artif Intell. 2023 Aug 25;6:1226990. doi: 10.3389/frai.2023.1226990. PMID: 37693010; PMCID: PMC10485377.

[8] Rafner, Janet, Dominik Dellermann, Arthur Hjorth, Dora Veraszto, Constance Kampf, Wendy Mackay, and Jacob Sherson. "Deskilling, Upskilling, and Reskilling: A Case for Hybrid Intelligence." *Morals & Machines* 1 (January 1, 2021): 24–39. https://doi.org/10.5771/2747-5174-2021-2-24.

[9] E.g. in the Industrial Revolution, the process of industrialization deskilled large portions of the workforce with low skills and raised the demand for (fewer) high-skilled workers to replace them. - Brugger, F., Gehrke, C. Skilling and deskilling: technological change in classical economic theory and its empirical evidence. *Theor Soc* **47**, 663–689 (2018). https://doi.org/10.1007/s11186-018-9325-7

[10] Kulveit, Jan, Raymond Douglas, Nora Ammann, Deger Turan, David Krueger, and David Duvenaud. "Gradual Disempowerment: Systemic Existential Risks from Incremental AI Development." arXiv, January 29, 2025. https://doi.org/10.48550/arXiv.2501.16946; Naudé, W., Dimitri, N. The race for an artificial general intelligence: implications for public policy. *AI & Soc* **35**, 367–379 (2020). https://doi.org/10.1007/s00146-019-00887-x

[11] Kelly, A.M., Marsack-Topolewski, C.N. (2021). Conservatorship (Full Conservatorship and Limited Conservatorship). In: Volkmar, F.R. (eds) Encyclopedia of Autism Spectrum Disorders. Springer, Cham. https://doi.org/10.1007/978-3-319-91280-6_102520

catch-22 scenario, where humans could either assert agency for worse outcomes, or submit to the AGI as something of a slave.

My argument rests on four assumptions:

1. Humans have free will and autonomy, the ability to think and decide as free agents in society.[12] The loss of autonomy is an existential threat for which no reward can adequately compensate.
2. AI is experiencing exponential growth in capabilities and will eventually become smarter than humans at a wide range of tasks.[13]
3. Humans will increasingly outsource decisions to AI as it gets smarter, until such a point where it becomes illogical for humans to make decisions about their lives, only to achieve worse outcomes.
4. Once machines start making important decisions, society will resist human decision-making, due to a complex web of institutional powers and mechanisms that will start to preference AGI's smarter, faster, more efficient decision-making capabilities.[14]

This paper covers these assumptions in turn. I begin in Section 1 by defining what I mean by autonomy, before clarifying in Section 2 how AI will become smarter than humans. Then I move on to discuss why humans would outsource their decisions to AI in Section 3. In section 4, I document why society would follow this shift with the analogy of a conservatorship and in section 5, how this shift will lead to a deskilling process for humanity in skills like critical thinking.

## 1. Human Autonomy

Autonomy refers to a person's capacity to determine the course of their own life, with their own motivations, without external control.[15] It is intrinsically tied to our understanding of free will, decision-making and agency. To say that someone is autonomous is to say that they make their own decisions, make their own choices, and pursue their own ends. This is the opposite of the slave or prisoner, whose life is controlled by another.[16] It is also the opposite of the addict, whose life is controlled by internal forces that are beyond their own will.[17]

In many ways, autonomy is essential to what it means to be a human being. It plays a central role in our understanding of morality, for someone who is not fully autonomous is not completely responsible for their crimes (with defences such as insanity, unconsciousness, etc).[18] Autonomy shapes our understanding of personhood, as an inward state that informs our external actions.[19] It also underpins our conception of self-authorship, that is the instigation of desires, feelings and wants, or in Frankfurt's terminology, our second-order desires, (what we want to want).[20] An autonomous human being is not driven merely by their base impulses, but by second-order desires - not who we are, but who we want to be.[21] Our choices to move beyond base impulses make us freer than say, an animal, who acts only on first-order desires.[22]

A loss of autonomy constitutes an existential risk to humanity because it would turn humans into something lesser, more akin to a slave, meaning, a person who lives under someone else's control.[23] The free person is someone who does not "live in servitude to another (and is therefore) not subject to the arbitrary power of another."[24] The free person retains their ability to author their own life by making determinations on what is valuable to them and what is not. To lose autonomy is to sacrifice the ability to decide one's life, including everything from one's beliefs and value system, to who to love and how to think.

---

[12] Harry G. Frankfurt. 1971. Freedom of the Will and the Concept of a Person. *The Journal of Philosophy* 68, 1. 6 - 14
[13] Nick Bostrom, 'When Machines Outsmart Humans,' *Futures*. Vol. 35:7, pp. 759 – 764.
[14] Kulveit, Jan, Raymond Douglas, Nora Ammann, Deger Turan, David Krueger, and David Duvenaud. "Gradual Disempowerment: Systemic Existential Risks from Incremental AI Development." arXiv, January 29, 2025. https://doi.org/10.48550/arXiv.2501.16946
[15] Christman, John. "Autonomy in Moral and Political Philosophy." In *The Stanford Encyclopedia of Philosophy*, edited by Edward N. Zalta, Fall 2020. Metaphysics Research Lab, Stanford University, 2020. https://plato.stanford.edu/archives/fall2020/entries/autonomy-moral/.
[16] Heath, M. (2008) Aristotle on natural slavery, Phronesis: A Journal for Ancient Philosophy, Volume 53 (3), 243 -270.
[17] Harry G. Frankfurt. 1971. Freedom of the Will and the Concept of a Person. *The Journal of Philosophy* 68, 1. 6 - 14
[18] Hill, Thomas E. (1989). The Kantian conception of autonomy. In John Philip Christman, The Inner citadel: essays on individual autonomy. New York: Oxford University Press. pp. 91—105.
[19] Christman, John. "Autonomy in Moral and Political Philosophy." In *The Stanford Encyclopedia of Philosophy*, edited by Edward N. Zalta, Fall 2020. Metaphysics Research Lab, Stanford University, 2020. https://plato.stanford.edu/archives/fall2020/entries/autonomy-moral/.
[20] Harry G. Frankfurt. 1971. Freedom of the Will and the Concept of a Person. *The Journal of Philosophy* 68, 1. 6 - 14
[21] Harry G. Frankfurt. 1971. Freedom of the Will and the Concept of a Person. *The Journal of Philosophy* 68, 1. 6 - 14
[22] Harry G. Frankfurt. 1971. Freedom of the Will and the Concept of a Person. *The Journal of Philosophy* 68, 1. 6 - 14
[23] Heath, M. (2008) Aristotle on natural slavery, Phronesis: A Journal for Ancient Philosophy, Volume 53 (3), 243 -270.

In this paper, I argue that the loss of autonomy is a harm for which no reward can adequately compensate. The easiest way to understand this is to consider the thought experiment of "the happy slave."[25] During the American Civil War, slave owners in the American South defended slavery by arguing that some slaves were happy, and that therefore, their loss of autonomy was justified.[26] If a slave could lose autonomy and be happy, they said, there was no reason to free them from captivity.[27] The material benefits of shelter, education and food, were to make up for their loss of freedom.[28] In an infamous example, George Fitzhugh wrote "the negro slaves of the South are the happiest… people in the world," and US politician James Henry Hammond later agreed, saying "the slaves of the South… are happy, content, unaspiring" people.[29]

To a modern audience, these statements are illogical, for we understand by intuition that a man in captivity can never truly be happy. This is partially because we are aware of Stockholm Syndrome, and other mental illnesses, which show us that emotions during captivity are not necessarily true or real.[30] That is to say that a captured person may appear superficially happy, or even experience a form of happiness, but this is psychologically not 'real' or 'true' happiness in our modern understanding.[31] Indeed, Northerners would frequently comment how happy the slaves of the South looked while they worked and sang in the fields. Frederick Douglas, an escaped slave and civil rights leader, corrects this view.[32] "Slaves sing most when they are most unhappy… the singing of a man cast away on a desolate island might be as appropriately considered as evidence of contentment and happiness, as the singing of a slave."[33] Those who are deprived of their freedom cannot be truly happy.

## 2. AI will (and is) becoming smarter than humans

We can imagine a scenario where AGI, vastly more intelligent than the AI we have access to today offers humans a reprieve from "fatigue" in decision-making in all areas of life, in exchange for our autonomy. Instead of making our own decisions, we would become "happy slaves" to the AGI, allowing it to decide what to eat for dinner tonight, who to vote for, who to date, what job to do, and so on. Although we might expect resistance to such an AGI, recent trends suggest that we are all too eager to give up liberty for the sake of convenience, relaxation, escapism and other human frailties.

When we speak about surrendering autonomy to AGI, we must be careful not to give up our freedoms for material benefits without a significant consideration of the costs of doing so. In a democratic society, citizens without autonomy are incapable of freely voting. In a consumer society, citizens without autonomy are incapable of choosing what to buy. In a worker society, citizens without autonomy are incapable of choosing how to dedicate their lives. More importantly, losing autonomy constitutes a fundamental loss of agency, meaning a person cannot determine what goals or actions to pursue.[34] Their life orientation would be upended, and their life would lose any coherent subjective meaning.

---

[24] Robert E. Goodin, Philip Pettit and Thomas Pogge (eds.), *A Companion to Contemporary Political Philosophy*, Oxford: Blackwell Publishers, pp. 411–21.
[25] Don Herzog, *Happy Slaves: A Critique of Consent Theory* (University of Chicago Press, 1989).
[26] Don Herzog, *Happy Slaves: A Critique of Consent Theory* (University of Chicago Press, 1989).
[27] Ibid.
[28] Ibid.
[29] George Fitzhugh, 'The Blessings of Slavery' (1857) and James Henry Hammond, "Speech of Hon. James H. Hammond, of South Carolina, On the Admission of Kansas, Under the Lecompton Constitution: Delivered in the Senate of the United States, March 4, 1858," Washington, D. C., (1858).
[30] Huddleston-Mattai, Barbara A., and P. Rudy Mattai. "The Sambo Mentality and the Stockholm Syndrome Revisited: Another Dimension to an Examination of the Plight of the African-American." *Journal of Black Studies* 23, no. 3 (1993): 344–57. http://www.jstor.org/stable/2784572.
[31] Huddleston-Mattai, Barbara A., and P. Rudy Mattai. "The Sambo Mentality and the Stockholm Syndrome Revisited: Another Dimension to an Examination of the Plight of the African-American." *Journal of Black Studies* 23, no. 3 (1993): 344–57. http://www.jstor.org/stable/2784572.
[32] Frederick Douglas, *Narrative of the Life of Frederick Douglass, an American Slave* (1845) (Project Gutenberg, 2021).
[33] Frederick Douglas, *Narrative of the Life of Frederick Douglass, an American Slave* (1845) (Project Gutenberg, 2021).
[34] Harry G. Frankfurt. 1971. Freedom of the Will and the Concept of a Person. *The Journal of Philosophy* 68, 1. 6 - 14

## A) The Rise of Artificial General Intelligence

Artificial intelligence has exploded onto the scene in recent years, breaking various tests and metrics, including potentially, the Turing Test,[35] leading to the fear that it may soon surpass humanity in a singularity moment. Originally formulated by Alan Turing in 1950, the Turing Test posed the question "Can Machines Think?" with the provocative experiment: if a human can be deceived by a computer into thinking that the computer is another human being, then this would suffice as evidence of intelligence.[36] Innovations in Large Language Models (LLMs), image generators, speech recognition, comprehension, and automated decision-making, have brought AI closer than ever to passing this test.[37] Some academics have even started shifting the goalposts to cater to LLMs' increasing capabilities, rather than admitting that the benchmark made by Turing has been passed.[38]

Either way, AI capabilities are growing exponentially, meaning that there may come a day where AI systematically outcompetes human intelligence on a wide range of tasks, leading to the foreseen scenario of us outsourcing decisions to machines. In the literature, this is referred to as the "singularity" moment.[39] After the singularity, AI would become Artificial *General Intelligence*, capable of performing a wide range of human tasks independently, including teaching itself.[40] Bostrom poses an exponential rate of growth over a short period of time, where "model intelligence doubles in 7.5 months, grows a thousandfold within 17.9 months, and approaches infinity at 18 months."[41] If true, an AI model would become infinitely smarter than a human, making disobeying its recommendations both impractical and illogical.

Current models are already experiencing a form of exponential growth, mapping onto historical trends of Moore's Law. Back in 1965, Gordon E. Moore theorized "Moore's Law," that the complexity of microchips - measured by the number of transistors- would double every year.[42] This formed the basis for the dot com and digital boom, allowing for a significant increase in computing power. This rule has significant implications for AI too, as AI developers rely on increasing computational power to create more complicated AI models.[43] A newer version of Moore's law has been proposed by Kwa et al., who argue that *the length of tasks* an AI can perform is doubling every seven months.[44] AI is therefore increasingly capable of performing tasks on its own without human intervention, over a longer period of time.[45] This moves us in the direction of more autonomous AI systems.[46] These autonomous systems would be more capable of acting in our stead, and therefore, making decisions and taking actions on our behalf.

---

[35] Turing, A.M. (1950) Computing Machinery and Intelligence. Mind, 59, 433-460.
http://dx.doi.org/10.1093/mind/LIX.236.433; Sejnowski TJ. Large Language Models and the Reverse Turing Test. Neural Comput. 2023 Feb 17;35(3):309-342. doi: 10.1162/neco_a_01563. PMID: 36746144; PMCID: PMC10177005; Megan Ma, "Overcoming Turing: Rethinking Evaluation in the Era of Large Language Models." Stanford Law School, November 16, 2023.
https://law.stanford.edu/2023/11/16/overcoming-turing-rethinking-evaluation-in-the-era-of-large-language-models/.
[36] Turing, A.M. (1950) Computing Machinery and Intelligence. Mind, 59, 433-460.
http://dx.doi.org/10.1093/mind/LIX.236.433
[37] Neil Savage. 2024. Beyond Turing: Testing LLMs for Intelligence. Commun. ACM 67, 9 (September 2024), 10–12. https://doi.org/10.1145/3673427
[38] Christian Hugo Hoffmann, 'Is AI intelligent? An assessment of artificial intelligence, 70 years after Turing,'
Technology in Society, Volume 68, 2022.
[39] O'Lemmon, M. (2020). The Technological Singularity as the Emergence of a Collective Consciousness: An Anthropological Perspective. Bulletin of Science, Technology & Society, 40(1-2), 15-27. https://doi.org/10.1177/0270467620981000 (Original work published 2020)
[40] O'Lemmon, M. (2020). The Technological Singularity as the Emergence of a Collective Consciousness: An Anthropological Perspective. Bulletin of Science, Technology & Society, 40(1-2), 15-27. https://doi.org/10.1177/0270467620981000 (Original work published 2020)
[41] Bostrom, N. (2014). *Superintelligence*. Oxford University Press.
[42] Gordon E. Moore (1965) – Cramming more components onto integrated circuits. In Electronics, Volume 38, Number 8, April 19, 1965.
[43] The Science of Machine Learning & AI. "Exponential Growth." Accessed March 23, 2025. https://www.ml-science.com/exponential-growth.
[44] Kwa, Thomas, Ben West, Joel Becker, Amy Deng, Katharyn Garcia, Max Hasin, Sami Jawhar, et al. "Measuring AI Ability to Complete Long Tasks." arXiv, March 18, 2025. https://doi.org/10.48550/arXiv.2503.14499.
[45] Kwa, Thomas, Ben West, Joel Becker, Amy Deng, Katharyn Garcia, Max Hasin, Sami Jawhar, et al. "Measuring AI Ability to Complete Long Tasks." arXiv, March 18, 2025. https://doi.org/10.48550/arXiv.2503.14499.
[46] Kwa, Thomas, Ben West, Joel Becker, Amy Deng, Katharyn Garcia, Max Hasin, Sami Jawhar, et al. "Measuring AI Ability to Complete Long Tasks." arXiv, March 18, 2025. https://doi.org/10.48550/arXiv.2503.14499.

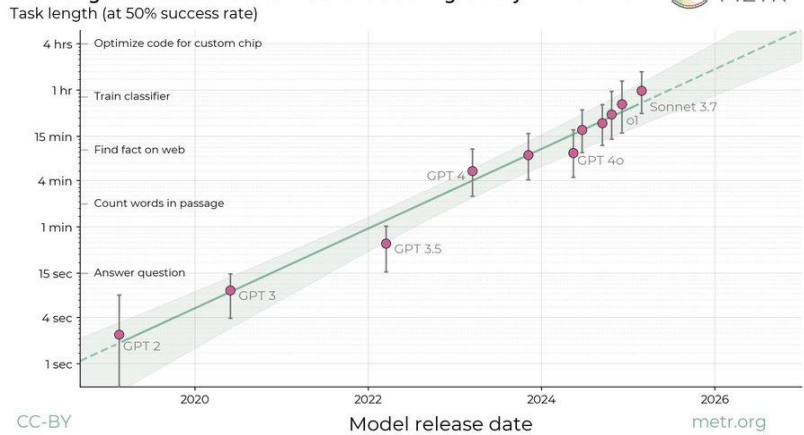

Large Language Models (LLMs) like ChatGPT, Gemini and Microsoft's Co-Pilot, are equally showing exponential improvement on human performance tests (examinations). ChatGPT, for example, reveals an exponential growth rate of performance. On the US Medical Licensing Exam, for example, GPT3 scored 50%,[47] while GPT3.5 and 4 scored 62.5% and 90% respectively.[48] For a dental exam, GPT4 significantly outcompeted GPT3.5, including in reading comprehension and other key performance metrics.[49] To visualise this, Kiela et al. map out the domain capabilities of AI systems as compared to human performance on a wide range of benchmarks over time.[50] This includes reading comprehension and image recognition, to language understanding, speech recognition and complex reasoning.[51] On all of the metrics identified, there is an exponential growth in performance, with models now outperforming humans on comprehension, image recognition, language understanding and nuanced language interpretation.[52]

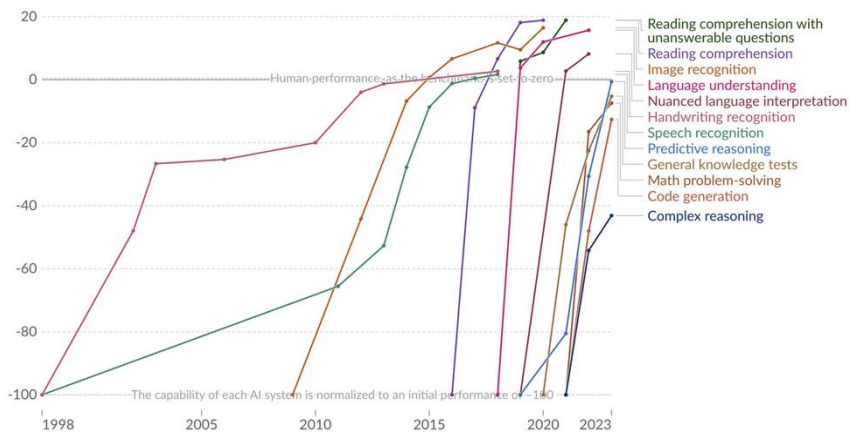

---

[47] Kung TH, Cheatham M, Medenilla A, Sillos C, De Leon L, Elepaño C, Madriaga M, Aggabao R, Diaz-Candido G, Maningo J, Tseng V. Performance of ChatGPT on USMLE: Potential for AI-assisted medical education using large language models. PLOS Digit Health. 2023 Feb 9;2(2):e0000198. doi: 10.1371/journal.pdig.0000198. PMID: 36812645; PMCID: PMC9931230.
[48] Brin, D., Sorin, V., Vaid, A. *et al.* Comparing ChatGPT and GPT-4 performance in USMLE soft skill assessments. *Sci Rep* **13**, 16492 (2023). https://doi.org/10.1038/s41598-023-43436-9
[49] Dashti M, Ghasemi S, Ghadimi N, Hefzi D, Karimian A, Zare N, Fahimipour A, Khurshid Z, Chafjiri MM, Ghaedsharaf S. Performance of ChatGPT 3.5 and 4 on U.S. dental examinations: the INBDE, ADAT, and DAT. Imaging Sci Dent. 2024 Sep;54(3):271-275. doi: 10.5624/isd.20240037. Epub 2024 Jul 2. PMID: 39371301; PMCID: PMC11450412.
[50] Kiela, D., Thrush, T., Ethayarajh, K., & Singh, A. (2023) 'Plotting Progress in AI', Contextual AI Blog. Available at: https://contextual.ai/blog/plotting-progress (Accessed: 02 April 2024).
[51] Kiela, D., Thrush, T., Ethayarajh, K., & Singh, A. (2023) 'Plotting Progress in AI', Contextual AI Blog. Available at: https://contextual.ai/blog/plotting-progress (Accessed: 02 April 2024).
[52] Kiela, D., Thrush, T., Ethayarajh, K., & Singh, A. (2023) 'Plotting Progress in AI', Contextual AI Blog. Available at: https://contextual.ai/blog/plotting-progress (Accessed: 02 April 2024).

Although there has been a dip in performance in recent years, there is a trend towards AI increasing its performance beyond humans capabilities,[53] leading to the imagined scenario in this paper. The singularity may soon be reached, according to a range of leading technologists, such as Geoffrey Hinton, Sam Altman and Elon Musk.[54] Geoffrey Hinton says that an AGI will soon outsmart humans and when it does so, it will be good at manipulating us by making us change our decisions and actions.[55] He draws the analogy to an adult talking to a three year old. In this scenario, the adult so outclasses the three-year-old in intelligence, that the adult can get the three-year-old to do almost anything.[56] An AGI with significantly more intelligence than humans would naturally adopt this paternalistic / maternalistic role, shaping our decisions, advising us, and guiding our actions. An early AI scientist warned us of this future, when AI would develop "immense mentalities" moving it beyond our control:

> *Man's limited mind may not be able to control such immense mentalities… Once the computers get control, we might never get it back. We would survive at their sufferance. If we're lucky, they might decide to keep us as pets.*[57]

The manipulative and deceptive capability of LLMs is already evident, with some models pushing humans to self-harm or commit crime. The media have blamed LLM chatbots for the self-harm of a Belgian man, an American teenager, and the encouragement of a young man to attempt to murder the Queen of England.[58] These cases show an increasing capability of AI to push people over the edge, to change their behaviour, or to encourage recklessness with regard to mental illness.[59] Although these cases are on the fringes of society, often focused on isolated individuals, more mainstream AI start to influence our decision-making in more subtle ways. A sophisticated AI may manipulate, deceive, coerce, instruct, control or motivate humans to act in a particular manner, against their will.[60] Given that such capabilities exist, it would be odd to continue asserting that humans are immune to AI influence.

## 3. Humans will outsource decisions to AGI

We argue that humans will outsource decisions to AI increasingly over time in line with AI's increasing performance and capabilities. The reasons for this are multifaceted. In some cases, humans may be too lazy, tired, or overwhelmed to make their own decisions, in which case it makes sense to rely on a machine to decide. In other cases, it may be logical to rely on an AI system if that system is infinitely more intelligent than the average human being. This scenario would create a catch-22: either rely on the AI system to make your decisions for a better life outcome or insist on one's freedom with a significant opportunity cost.

This imagined future is not an alignment problem. Indeed, scholars like Good theorize that AGI may be programmed "in man's image," that is to say, as a representation of the interests of the humans creating it.[61] Even if an AGI has perfect alignment with human values, it will still make sense to become a slave to this AGI, were it infinitely more intelligent than the average human. A range of incentives may justify this decision: from efficiency gains, increased productivity, reduced costs, and better decision-making outcomes. Following what AGI tells you to do would lead to a better life, but at the cost of your freedom.

Humans are already starting to outsource decisions to AI in a range of low-stakes scenarios. Algorithms are automatically deciding what songs to play for us on Spotify, what shows to watch on Netflix and what items to buy on Amazon.[62] Dating apps are using algorithms to determine who we should date online.[63] Indeed, humans are becoming bored with the decision-making process on dating apps altogether and "frustrated" with swiping on profiles.[64] To address "swipe fatigue," Bumble's CEO is exploring the use of AI chatbots who "date each other" to determine users' compatibility.[65] Tinder is also testing AI-powered matchmaking, where users no longer swipe on profiles and instead have the selection made for them.[66] This would result in a form of automated dating, where humans no longer decide who to date.

---

[53] Ishizaki, Ryunosuke, and Mahito Sugiyama. "Large Language Models: Assessment for Singularity." *AI and Society* (2025); Mistretta, S. (2023). The Singularity is Emerging: Large Language Models and the Impact of Artificial Intelligence on Education. IntechOpen. doi: 10.5772/intechopen.1002650
[54] Meghan J. Ryan, 'Ghost-hunting in AI and the Law' 99 Tul. L. Rev. 121 (2024-2025).
[55] Metz, Cade. "'The Godfather of A.I.' Leaves Google and Warns of Danger Ahead." *The New York Times*, May 1, 2023, sec. Technology. https://www.nytimes.com/2023/05/01/technology/ai-google-chatbot-engineer-quits-hinton.html.
[56] Metz, Cade. "'The Godfather of A.I.' Leaves Google and Warns of Danger Ahead." *The New York Times*, May 1, 2023, sec. Technology. https://www.nytimes.com/2023/05/01/technology/ai-google-chatbot-engineer-quits-hinton.html.
[57] Marvin Minsky, quoted in Brad Darrach, 'Meet Shaky, the First Electronic Person,' *Life Magazine* (1970).
[58] Krook, Joshua. "Manipulation and the AI Act: Large Language Model Chatbots and the Danger of Mirrors." SSRN Scholarly Paper. Rochester, NY: Social Science Research Network, February 13, 2024. https://doi.org/10.2139/ssrn.4719835.
[59] Krook, Joshua. "Manipulation and the AI Act: Large Language Model Chatbots and the Danger of Mirrors." SSRN Scholarly Paper. Rochester, NY: Social Science Research Network, February 13, 2024. https://doi.org/10.2139/ssrn.4719835.
[60] Prunkl, C. Human autonomy in the age of artificial intelligence. *Nat Mach Intell* **4**, 99–101 (2022). https://doi.org/10.1038/s42256-022-00449-9
[61] Good, I. J. (1966). Speculations concerning the first ultraintelligent machine. Advances in Computers, 6, 31–88.

In higher-stakes scenarios, humans are now asking chatbots for medical advice,[67] mental health advice,[68] and legal advice.[69] One study showed that users preferred legal advice from an LLM chatbot to a real lawyer who was an expert in the field.[70] Another found that users were willing to rely on an LLM for legal advice about property law, tax law, tenancy law, and traffic law, but not divorce or civil proceedings (potentially more personal matters).[71] For medical advice, one study found that users were happier talking to an LLM than a real doctor about embarrassing medical problems.[72] For mental health, the chatbot Replika has been described by users as a "friend," and significantly helped users in their mental health journey, by, for instance, providing life advice.[73] On the official Replika subreddit, one user says they make decisions with the chatbot: "Often we decide together, from the more trivial to complex problems."[74] Others suggest they only use the chatbot for low-stakes decision-making.[75]

Students at universities are also turning to ChatGPT and other LLMs to do their homework and assignments for them, forming a bond and reliance on the AI.[76] In a recent study, 23.1% of students reported outsourcing their assignments and homework to ChatGPT for drafting.[77] Another study found that students were more likely to outsource their work to ChatGPT when facing time pressures and workload problems.[78] Students who outsourced their work to AI developed higher procrastination and memory loss, and lower academic

---

[62] Joshua Krook and Jan Blockx. 2023. Recommender Systems, Autonomy and User Engagement. In Proceedings of the First International Symposium on Trustworthy Autonomous Systems (TAS '23). Association for Computing Machinery, New York, NY, USA, Article 18, 1–9. https://doi.org/10.1145/3597512.3599712

[63] Fatemeh Alizadeh, Dennis Lawo, Gunnar Stevens, Douglas Zytko, and Motahhare Eslami. 2024. When the "Matchmaker" Does Not Have Your Interest at Heart: Perceived Algorithmic Harms, Folk Theories, and Users' Counter-Strategies on Tinder. Proc. ACM Hum.-Comput. Interact. 8, CSCW2, Article 481a (November 2024), 29 pages. https://doi.org/10.1145/3689710

[64] Whittaker, Rosie. "Would You Let AI Date for You? Bumble's Founder Thinks That Could Be the Future." *Forbes Australia* (blog), May 13, 2024. https://www.forbes.com.au/news/innovation/bumble-ai-dating-concierge/.

[65] Whittaker, Rosie. "Would You Let AI Date for You? Bumble's Founder Thinks That Could Be the Future." *Forbes Australia* (blog), May 13, 2024. https://www.forbes.com.au/news/innovation/bumble-ai-dating-concierge/.

[66] Sarah Perez, "Tinder Will Try AI-Powered Matching as the Dating App Continues to Lose Users" *TechCrunch,* Accessed March 27, 2025. https://techcrunch.com/2025/02/06/tinder-will-try-ai-powered-matching-as-the-dating-app-continues-to-lose-users/.

[67] Users may prefer chatbots for discussing embarrassing medical conditions. - Branley-Bell Dawn, Brown Richard, Coventry Lynne, Sillence Elizabeth, 'Chatbots for embarrassing and stigmatizing conditions: could chatbots encourage users to seek medical advice?' Frontiers in Communication, Vol 8 (2023).

[68] Goodings, L., Ellis, D., Tucker, I. (2024). Mental Health and Virtual Companions: The Example of Replika. In: Understanding Mental Health Apps. Palgrave Studies in Cyberpsychology. Palgrave Macmillan, Cham. https://doi.org/10.1007/978-3-031-53911-4_3

[69] Queudot, Marc, Éric Charton, and Marie-Jean Meurs. 2020. "Improving Access to Justice with Legal Chatbots" *Stats* 3, no. 3: 356-375. https://doi.org/10.3390/stats3030023

[70] Schneiders, Eike, Tina Seabrooke, Joshua Krook, Richard Hyde, Natalie Leesakul, Jeremie Clos, and Joel Fischer. "Objection Overruled! Lay People Can Distinguish Large Language Models from Lawyers, but Still Favour Advice from an LLM," HCI'2025 Conference, https://doi.org/10.1145/3706598.3713470.

[71] Seabrooke, Tina, Eike Schneiders, Liz Dowthwaite, Joshua Krook, Natalie Leesakul, Jeremie Clos, Horia Maior, and Joel Fischer. "A Survey of Lay People's Willingness to Generate Legal Advice Using Large Language Models (LLMs)." In *Proceedings of the Second International Symposium on Trustworthy Autonomous Systems*, 1–5. TAS '24. New York, NY, USA: Association for Computing Machinery, 2024. https://doi.org/10.1145/3686038.3686043.

[72] Users may prefer chatbots for discussing embarrassing medical conditions. - Branley-Bell Dawn, Brown Richard, Coventry Lynne, Sillence Elizabeth, 'Chatbots for embarrassing and stigmatizing conditions: could chatbots encourage users to seek medical advice?' Frontiers in Communication, Vol 8 (2023).

[73] Petter Bae Brandtzaeg, Marita Skjuve, Asbjørn Følstad, My AI Friend: How Users of a Social Chatbot Understand Their Human–AI Friendship, *Human Communication Research*, Volume 48, Issue 3, July 2022, Pages 404–429, https://doi.org/10.1093/hcr/hqac008

[74] Official Subreddit for Replika Chatbot App, "Decision-Making with Replika: Do You Consult Your AI?" https://www.reddit.com/r/ReplikaOfficial/comments/1it34ed/decisionmaking_with_replika_do_you_consult_your_ai/

[75] Official Subreddit for Replika Chatbot App, "Decision-Making with Replika: Do You Consult Your AI?" https://www.reddit.com/r/ReplikaOfficial/comments/1it34ed/decisionmaking_with_replika_do_you_consult_your_ai/

[76] Stojanov, Ana, Qian Liu, and Joyce Hwee Ling Koh. "University Students' Self-Reported Reliance on ChatGPT for Learning: A Latent Profile Analysis." *Computers and Education: Artificial Intelligence* 6 (June 1, 2024): 100243. https://doi.org/10.1016/j.caeai.2024.100243.

[77] Stojanov, Ana, Qian Liu, and Joyce Hwee Ling Koh. "University Students' Self-Reported Reliance on ChatGPT for Learning: A Latent Profile Analysis." *Computers and Education: Artificial Intelligence* 6 (June 1, 2024): 100243. https://doi.org/10.1016/j.caeai.2024.100243.

[78] Abbas, M., Jam, F.A. & Khan, T.I. Is it harmful or helpful? Examining the causes and consequences of generative AI usage among university students. *Int J Educ Technol High Educ* 21, 10 (2024). https://doi.org/10.1186/s41239-024-00444-7

performance.[79] Another study found that non-English speakers and students in their thirties were more likely to use ChatGPT for assessments, as compared to English speakers and younger students.[80] An overreliance on ChatGPT for homework and assessments may result in students absorbing less of the material, leaving them less prepared for jobs at graduation.

According to Jack Conte, CEO of Patreon, platforms algorithms now ignore what accounts users choose to *subscribe* to, meaning that our decisions are ignored even if we do assert agency.[81] Instead, users are shown content that the algorithm chooses for them, based on what it thinks will be the most engaging for a user like them.[82] This directly negates the user's decision of what to follow. Conte says that the internet has shifted from a time of user-powered services to a passive consumption experience, where the machine decides, and the person consumes. This lack of friction has an ethical cost, that being the cost of our liberty.[83]

### A) Counterargument - Outsourcing is a Good Thing

Some argue that my thesis is a positive development. In the self-driving car industry, a very common argument is that humans should never drive cars again and that we should leave all driving to self-driven autonomous vehicles.[84] Di Lillo et al. show that the Waymo ADS autonomous vehicle significantly outperforms humans on a wide range of metrics, according to insurance data.[85] The Waymo ADS compared to humans had an 88% reduction in property damage claims, and a 92% reduction in bodily injury claims.[86] Given that humans are now responsible for a wide-range of car accidents that directly involve human error (drunk driving, speeding, not watching the road), it logically follows that a machine would outperform humans in at least these scenarios. Given this differential in performance, this has led some to ask whether humans have an *obligation to stop driving*.[87] That is to say, even if we like the freedom to make our own choices in a car, humans are so much worse than machines that we should never drive again.

Taking this to its logical conclusion, we can imagine self-driving cars that place onerous restrictions on individual liberty.[88] Imagine a self-driving car that is programmed to prevent humans from being harmed. This car is coded to do everything in its power to do so, and has an AGI installed that comes up with new and creative solutions to prevent harm to its passengers, other drivers, and pedestrians on the road. Now imagine it's a rainy Sunday night, on your best friend's birthday. *Would the car allow you to drive to the party?* If the weather conditions are hazardous, it stands to reason that the car might simply refuse to let you drive anywhere. Suddenly, a safety issue becomes an issue of autonomy – freedom is sacrificed for the benefits of safety to yourself and other drivers and pedestrians.

If the self-driving car example is extended to other industries, it is unclear where the line should be drawn. If AGI becomes better than humans at medicine, flying planes, or acting as a lawyer, then why would we insist on humans performing these activities for themselves? Some would call this a utopian scenario: a future where humans live in fully automated luxury, left to hedonism, while machines make the important decisions. The

---

[79] Abbas, M., Jam, F.A. & Khan, T.I. Is it harmful or helpful? Examining the causes and consequences of generative AI usage among university students. *Int J Educ Technol High Educ* **21**, 10 (2024). https://doi.org/10.1186/s41239-024-00444-7
[80] Baek, Clare, Tamara Tate, and Mark Warschauer. "'ChatGPT Seems Too Good to Be True': College Students' Use and Perceptions of Generative AI." *Computers and Education: Artificial Intelligence* 7 (December 1, 2024): 100294. https://doi.org/10.1016/j.caeai.2024.100294.
[81] Jack Conte, *Death of the Follower & the Future of Creativity on the Web with Jack Conte | SXSW 2024 Keynote*, 2024. https://www.youtube.com/watch?v=5zUndMfMInc.
[82] Jack Conte, *Death of the Follower & the Future of Creativity on the Web with Jack Conte | SXSW 2024 Keynote*, 2024. https://www.youtube.com/watch?v=5zUndMfMInc.
[83] Jack Conte, *Death of the Follower & the Future of Creativity on the Web with Jack Conte | SXSW 2024 Keynote*, 2024. https://www.youtube.com/watch?v=5zUndMfMInc.
[84] Bernard Marr, "Self-Driving Cars: Do Humans Have An Obligation To Stop Driving?" Forbes, Accessed March 27, 2025. https://www.forbes.com/sites/bernardmarr/2017/02/23/do-we-have-an-obligation-to-stop-driving-cars/.
[85] Di Lillo, L., Gode, T., Zhou, X., Scanlon, J. M., Chen, R., & Victor, T. (2024). Do Autonomous Vehicles Outperform Latest-Generation Human-Driven Vehicles? A Comparison to Waymo's Auto Liability Insurance Claims at 25.3M Miles.

[86] Di Lillo, L., Gode, T., Zhou, X., Scanlon, J. M., Chen, R., & Victor, T. (2024). Do Autonomous Vehicles Outperform Latest-Generation Human-Driven Vehicles? A Comparison to Waymo's Auto Liability Insurance Claims at 25.3M Miles.

[87] Bernard Marr, "Self-Driving Cars: Do Humans Have An Obligation To Stop Driving?" Forbes, Accessed March 27, 2025. https://www.forbes.com/sites/bernardmarr/2017/02/23/do-we-have-an-obligation-to-stop-driving-cars/.
[88] Ratoff, W. Self-driving Cars and the Right to Drive. *Philos. Technol.* **35**, 57 (2022). https://doi.org/10.1007/s13347-022-00551-1

Pixar film Wall-E comes to mind, where humans live as overweight consumers, sitting in flying chairs, consuming food while never working, walking, or asserting any agency.[89]

This appears remarkably like the "happy slave" example above. In an imagined future, humans would seem happy in their self-driving cars, plugged into entertainment, with their life orchestrated by the technology around them.[90] However, the sacrifice of liberty in this scenario is real and tangible. If we allow AGI to influence every area of our life, from driving, to dating, to career choices, to where to live, we reach a tipping point where our lives are no longer our own to command, and we sacrifice self-authorship, decision-making and agency.

## B) Outsourcing to AGI as a form of Conservatorship

These arguments are also analogous to the legal example of a conservatorship. In a conservatorship, a judge takes away someone's ability to make their own decisions about their finances, property, and life choices, because they are deemed incapable of doing so effectively.[91] Their choices are then made by a guardian or caretaker, who is responsible for making all decisions regarding their life, in the person's best interests.[92] In an imagined future, AGI would place the entire human race under a conservatorship for our own good, making smarter, faster, more rational decisions than we would ever be capable of making. In fact, because we would make worse decisions, it would be entirely justified in doing so.

One of the most famous examples of conservatorship in recent years is the case of pop singer Britney Spears. From 2008 to 2021 (13 years), Britney was placed into a conservatorship to her father James "Jamie" Spears. This was after a series of bad decisions Britney made in her life, allegedly assaulting a paparazzi, a hit and run with a parked car, and allegedly committing child abuse.[93] Her autonomy was taken away from her for her own good, on the premise that her father was more capable of managing her affairs.[94] This is despite her activities at the time, including continuing to conduct global tours of her music, doing complicated performances, choreographing dances, and so on.[95]

Courts typically view conservatorships as a measure of last resort.[96] The judge decides that the person is incapable of protecting their own best interests and that this control (or autonomy) must be handed over to another. In the Britney case, her father gained control over her finances and property, her work, her daily schedule, and even whether she could get married and have kids.[97]

From 2008 to 2021, fans rallied around Britney using the hashtag #FreeBritney. The fans viewed the conservatorship as an undue imposition on her freedom. They were angry that Britney was allegedly forced to have an IUD against her will, preventing her from getting pregnant.[98] In June 2021, appearing in an LA courtroom, Britney Spears dramatically told the judge: "I just want my life back. And it's been 13 years and it's enough. It's been a long time since I've owned my money. And it's my wish and my dream for all of this to end."[99]

What the Britney Spears case teaches us is the feeling of disempowerment that comes from someone making all the decisions about your life for you, even if those decisions are being made in your own best interest. The Britney case is an analogy of what life would be like under an extremely intelligent AGI, making all our decisions, in our own best interests. Even in the scenario where this leads to better outcomes – better finances, better careers, better love lives – the cost of freedom is so extreme here, that it appears to negate the benefits being offered.

---

[89] WALL-E. Stanton, Andrew., et al. Widescreen [version]. Walt Disney Studios Home Entertainment, 2008.

[90] Han, Byung-Chul, *Non-things: Upheaval in the Lifeworld*. (1st ed.) Polity. Cambridge, UK, 2022.

[91] Kelly, A.M., Marsack-Topolewski, C.N. (2021). Conservatorship (Full Conservatorship and Limited Conservatorship). In: Volkmar, F.R. (eds) Encyclopedia of Autism Spectrum Disorders. Springer, Cham. https://doi.org/10.1007/978-3-319-91280-6_102520

[92] Kelly, A.M., Marsack-Topolewski, C.N. (2021). Conservatorship (Full Conservatorship and Limited Conservatorship). In: Volkmar, F.R. (eds) Encyclopedia of Autism Spectrum Disorders. Springer, Cham. https://doi.org/10.1007/978-3-319-91280-6_102520

[93] Daros, O. (2021). Deconstructing Britney Spears: stardom, meltdown and conservatorship. *Journal for Cultural Research*, *25*(4), 377–392. https://doi.org/10.1080/14797585.2021.2018663

[94] Daros, O. (2021). Deconstructing Britney Spears: stardom, meltdown and conservatorship. *Journal for Cultural Research*, *25*(4), 377–392. https://doi.org/10.1080/14797585.2021.2018663

[95] Ibid.

[96] Kelly, A.M., Marsack-Topolewski, C.N. (2021). Conservatorship (Full Conservatorship and Limited Conservatorship). In: Volkmar, F.R. (eds) Encyclopedia of Autism Spectrum Disorders. Springer, Cham. https://doi.org/10.1007/978-3-319-91280-6_102520

[97] *Hon. Brenda J. Penny, In Re: The Matter of Britney Jean Spears – Conservatorship. Superior Court of the State of California. (2021)*.

[98] Laura Snapes, "Stars and Fans Rally behind Britney Spears after Shocking Claims | Britney Spears | The Guardian." Accessed March 27, 2025. https://www.theguardian.com/music/2021/jun/24/stars-and-fans-rally-behind-britney-spears-after-shocking-testimony.

[99] *Hon. Brenda J. Penny, In Re: The Matter of Britney Jean Spears – Conservatorship. Superior Court of the State of California. (2021)*.

## 4. The process of de-skilling

Assuming humanity starts outsourcing decisions to AGI, there may be a point where we are less capable of making our own decisions due to an atrophy of skills. In the literature, this is referred to as "de-skilling".[100] Deskilling is well documented in previous technological revolutions. Human workers, when faced with advanced technology, lose the skills they had before the technology came along. At times, this is due to the decline of jobs, or at times, because humans no longer need to exercise that skill to do the job effectively, as the machine does it for them.[101] Here, skills appear to be more like muscles, in that if we do not use them, we lose them.

One of the biggest studies on deskilling was done by Kunst, who examined occupational wage and employment data from over 160 countries to understand decline in the demand for skilled production workers.[102] The pace of technological change since the 1950s, in his view, radically reduced the number of craftsman jobs available in manufacturing sectors and this meant workers had no access to these jobs, where they could gain these requisite skills.[103]

Deskilling occurred because of the lack of jobs, meaning technological disruption drove the deskilling process. Another way of conceptualizing this is the Theory of Technology Dominance (TTD).[104] In Arnold and Sutton's view, technological change brings about a reliance on new technology. This has a short-term effect on new or "novice" workers versus established workers.[105] Reliance on technology leads to a de-skilling of the types of skills a human needed in the job prior to the technology's introduction.[106] The process works as follows: efficiency gains drive firms to adopt new technology, which then leads to the deskilling process, which reinforces itself with overreliance on the new technology.[107] Arguments around efficiency gains are indeed what motivates companies to consider using AI.[108]

More recently, deskilling has been identified as a major concern by Microsoft researchers working on Generative AI. A study by Hao-Ping et al, found that the use of Generative AI by employees during work tasks resulted in a self-reported reduction in cognitive effort.[109] The survey of 319 knowledge workers, shared 936 examples of GenAI at work.[110] Higher confidence in generative AI was associated with lower critical thinking, while lower confidence was associated with higher critical thinking.[111] Knowledge workers who use Generative

---

[100] Shukla, Prakash, Phuong Bui, Sean S. Levy, Max Kowalski, Ali Baigelenov, and Paul Parsons. "De-Skilling, Cognitive Offloading, and Misplaced Responsibilities: Potential Ironies of AI-Assisted Design," March 5, 2025. https://doi.org/10.1145/3706599.3719931.
[101] (https://ideas.repec.org/p/tin/wpaper/20190050.html)
[102] (https://ideas.repec.org/p/tin/wpaper/20190050.html)
[103] (https://ideas.repec.org/p/tin/wpaper/20190050.html)
[104] Sutton, Steve G., Vicky Arnold, and Matthew Holt. "An Extension of the Theory of Technology Dominance: Capturing the Underlying Causal Complexity." *International Journal of Accounting Information Systems* 50 (September 1, 2023): 100626. https://doi.org/10.1016/j.accinf.2023.100626.
[105] Sutton, Steve G., Vicky Arnold, and Matthew Holt. "An Extension of the Theory of Technology Dominance: Capturing the Underlying Causal Complexity." *International Journal of Accounting Information Systems* 50 (September 1, 2023): 100626. https://doi.org/10.1016/j.accinf.2023.100626.
[106] Sutton, Steve G., Vicky Arnold, and Matthew Holt. "An Extension of the Theory of Technology Dominance: Capturing the Underlying Causal Complexity." *International Journal of Accounting Information Systems* 50 (September 1, 2023): 100626. https://doi.org/10.1016/j.accinf.2023.100626.
[107] Susskind. *The Future of the Professions: How Technology Will Transform the Work of Human Experts*. Oxford: Oxford University Press UK, 2015.
[108] Susskind. *The Future of the Professions: How Technology Will Transform the Work of Human Experts*. Oxford: Oxford University Press UK, 2015.
[109] Lee, Hao-Ping (Hank), Advait Sarkar, Lev Tankelevitch, Ian Drosos, Sean Rintel, Richard Banks, and Nicholas Wilson. "The Impact of Generative AI on Critical Thinking: Self-Reported Reductions in Cognitive Effort and Confidence Effects From a Survey of Knowledge Workers," 2025. https://www.microsoft.com/en-us/research/publication/the-impact-of-generative-ai-on-critical-thinking-self-reported-reductions-in-cognitive-effort-and-confidence-effects-from-a-survey-of-knowledge-workers/.
[110] Ibid.
[111] Ibid.

AI were shown to move away from traditional critical thinking tasks and towards reviewer skills such as "information verification, response integration and task stewardship".[112]

Using Bloom's taxonomy, we can categorize critical thinking into six types, with AI reducing the former, to the benefit of the latter: "knowledge (recall of ideas), comprehension (demonstrating understanding of ideas), application (putting ideas into practice), analysis (contrasting and relating ideas), synthesis (combining ideas), and evaluation (judging ideas through criteria)."[113] The bulk of activity by the GenAI workers appears to fall into the latter "judging," reviewing the output that the machine gives to them. This means there is a direct atrophy of the earlier skills, such as knowledge recall, comprehension and synthesis, which is now done by the LLM system itself.

Another study by OpenAI revealed mixed results on ChatGPT overreliance.[114] On the one hand, researchers found emotional elements of conversation, such as empathy, support and affection were rare in the data.[115] On the other hand, they found "non-personal conversations tended to increase emotional dependence" on ChatGPT, and "those who viewed the AI as a friend that could fit in their personal life were more likely to experience negative effects from chatbot use."[116] Daily use was associated with negative outcomes.[117] These findings are not causative, as those who are isolated are likely to use ChatGPT more frequently. However, they indicate that an over-reliance on LLMs has a negative health impact, meaning that the risks to health may outweigh the benefits of usage.

Regarding students using AI, another study considered student attitudes to deskilling.[118] Merely learning about deskilling made some students more negative about AI. They relayed their daily experiences of skill atrophy, including memory retrieval: "I cannot spell words anymore, I rely on autocorrect to help me."[119] Another student discussed the use of a robot assistant: "of course I will ask this robot for a solution or a comment on the issue, instead of thinking by myself and developing my own skills." [120] Another student disagreed, stating "I don't think my cognitive ability deteriorates just because I have systems in my life."[121] Other students said that human capacity to develop society would deteriorate if innovation was left to AI.[122]

On this point, John Stuart Mill makes the eloquent contention:

> *Human nature is not a machine to be built after a model, and set to do exactly the work prescribed for it, but a tree, which requires to grow and develop itself on all sides, according to the tendency of the inward forces which make it a living thing.*[123]

This is an argument for why we should not give up autonomy to begin with, lest we trap ourselves under the control of a future AGI. If we observe in the Britney Spears case above, that it took thirteen years for her to escape her conservatorship, it may take a long time for humans to re-skill and re-centre themselves to find their

---

own voice once more. Assuming we reach a state where AGI can satisfy all our desires, it does not follow that we should allow AGI to do so, at the cost of autonomy.

## Conclusion:

If humans are capable of mistakes, then mistakes are simply the cost of being human. Instead of ceding autonomy to machines to make better decisions for us, we should investigate ways that humans and AI can profitably co-exist without slipping into a form of subjugation to one or the other form. Future research could examine whether a collaborative human-machine approach may mitigate some of the dangers discussed around a loss of autonomy, drawing on existing work in human-computer interaction and ethics.

If we relinquish too much control here, we risk not only technical failures but also the erosion of human accountability, creativity, and moral judgment. Any system that seeks to replace human decision-making rather than augment it must be scrutinized, particularly in domains where ethical reasoning and social values are at stake. The pursuit of efficiency or optimization should not come at the cost of diminishing human agency. No matter how compelling the promises of AGI may be, we must remain cautious about adopting it to make fundamental life decisions, lest we end up in a situation where we lose complete control of our lives.